# The chemistry of extra-solar materials from white dwarf planetary systems

Siyi Xu (许偲艺),[1] Laura K. Rogers,[2] and Simon Blouin[3]

[1]*Gemini Observatory/NSF's NOIRLab, 670 N. A'ohoku Place, Hilo, Hawaii, 96720, USA*
[2]*Institute of Astronomy, University of Cambridge, Madingley Road, Cambridge CB3 0HA, UK*
[3]*Department of Physics and Astronomy, University of Victoria, Victoria, BC V8W 2Y2, Canada*

## 1. INTRODUCTION

The search for life on other planets is one of the grand scientific goals of our time. The Decadal Survey on Astronomy and Astrophysics 2020 (Astro2020)[1] sets the top priority in the coming decade to be the pursuit of a new space observatory called the Habitable World Observatory (WHO), with the goal to directly detect Earth-like planets, characterize its atmosphere, and search for biosignatures. However, whether a gas is a biosignature depends on the geochemical cycles (Shahar et al. 2019). We can't fully identify a gas as a biosignature unless we know the planet's (abiotic) geochemical cycles, and that means having an inventory of the elements within that planet.

Chemical inventories of exoplanets are difficult to obtain due to the limitations of observational techniques. To directly constrain the composition of planets, accurate measurements of their masses and radii are required. The transit technique measures the flux change of a star as a planet transits the stellar disk. This flux change is proportional to $(R_P/R_*)^2$, the square of the ratio of the planet's radius $R_P$ to the star's radius $R_*$; if the radius of the star is known then the radius of the planet can be inferred. The transit technique can only detect planets whose orbits are close to edge on (when the inclination of the orbit is close to 90degrees). For example, an Earth-like planet that transits a Sun-like star would produce a transit depth of 0.008% – such a precision has yet to be achieved from the ground due to the variable atmospheric condition and different telescope/instrument systematics. Space telescopes are much better and the *Kepler* Mission has reached a precision of 0.002% in the best-case scenario (Gilliland et al. 2011).

Another popular technique to detect planets around stars is the radial velocity technique, which measures the wobble of a star due to the gravitational interaction with an orbiting planet. This technique provides a minimum planetary mass ($M_P \sin i$), where $M_P$ is the mass of the planet and $i$ is the orbital inclination. For example, the presence of Earth affects the radial velocity of the Sun, but only by 10 cm s$^{-1}$. That is still beyond the reach of current Extreme Precision Radial Velocity (EPRV) Spectrographs, which can achieve a precision of ≈ 30 cm s$^{-1}$ under the best circumstances (e.g., Pepe et al. 2021; Seifahrt et al. 2022). Future instrumentation on the *Extremely Large Telescopes* (ELTs) will likely have even better precision. However, stellar activity will remain a major challenge in detecting radial velocity signals from an Earth-like planets (e.g., Wright 2018).

For planets which are detected via both the transit technique and the radial velocity technique, the planetary mass can be derived because the inclination is known ($i ≈$ 90degrees). Using the mass and radius measurements the bulk density is inferred, $\rho_{\rm bulk} = M_P/(4/3\pi R_P^3)$ (also see discussion in the chapter Transiting Exoplanet Atmospheres in the Era of JWST). The first planet for which this was done was HD209458. The bulk density was found to be low (0.38 g cm$^{-3}$) and given the large radius (1.27 Jupiter radii), it was designated a gas giant (Charbonneau et al. 2000). Planet interior models provide theoretical mass-radius relations for different compositions and interior structures. For example, the planetary models may assume a single composition, e.g., 100% iron, or 100% hydrogen/helium in gaseous form; or multi-layered models with different compositions per layer, e.g., metallic iron core, $MgSiO_3$ mantle, water, or more exotic compositions (e.g., Seager et al. 2007; Dorn et al. 2015). The difficulty with this analysis is that the models are degenerate, with vastly different compositions and structures producing similar mass-radius curves. Therefore, to study the interior composition and structure, further techniques are required.

White dwarf planetary systems provide a unique way to measure the bulk composition of exoplanetary material. As introduced in the previous chapter "The evolution and delivery of rocky extra-solar materials to white dwarfs", extrasolar asteroids/comets/moons which have survived the evolution of their host star can end up in the atmosphere of the white dwarf. Asteroids and boulders appear to be the most common pollutants (see previous chapter) and in this chapter, we use the term "asteroids" to refer to the parent body that is polluting the atmosphere. The presence of the planetary material is detected via absorption lines of heavy elements[2]. White dwarfs with these absorption features are called 'polluted' white dwarfs. Polluted white dwarfs were expected to be rare objects because white dwarfs have high surface gravities, therefore, these heavy elements will settle out of the white dwarf's atmospheres in a short amount of time (Paquette et al. 1986), as illustrated in Figure 1. However, high-resolution spectroscopic surveys found that 25 – 50% of white dwarfs are polluted (Zuckerman et al. 2003, 2010; Koester et al. 2014). The mechanism responsible for making a polluted white dwarf must be common and efficient. In the early days, accretion from the interstellar medium was proposed as the source of the pollution, but this idea was rejected due to (i) the small amount of carbon pollution, which is a major element in the interstellar medium (Jura 2006), and (ii) the lack of correlation between the locations of the polluted white dwarfs and the interstellar clouds (Farihi et al. 2010).

---

[1] https://nap.nationalacademies.org/catalog/26141/pathways-to-discovery-in-astronomy-and-astrophysics-for-the-2020s

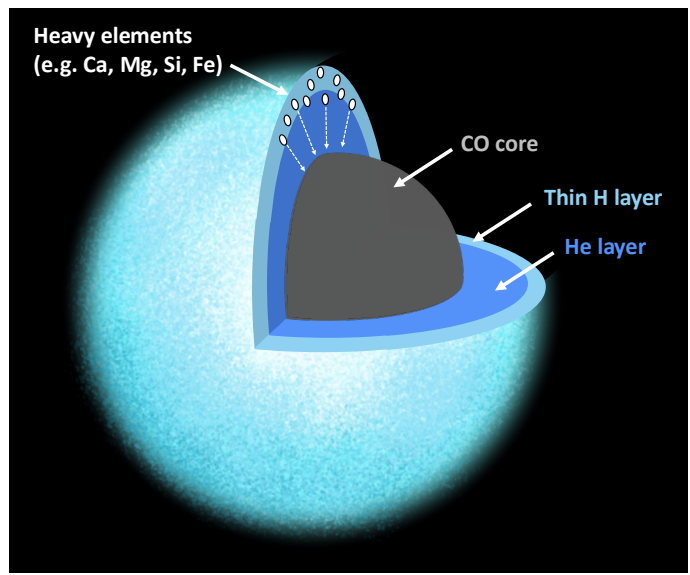

**Figure 1.** Cross section of a polluted white dwarf, where heavy elements (e.g., Ca, Mg, Si, Fe) arrive in the atmosphere of the white dwarf and sink down over time. Most white dwarfs have a carbon/oxygen-rich core, but only the outer hydrogen/helium layer is directly observable. This figure is for illustration purpose and the size of the different layers is not to scale.

There is strong theoretical and observational evidence that white dwarfs are accreting from planetary material. There are different mechanisms that can deliver exoplanetary material into the Roche lobe of the white dwarf, as discussed in the section "Stage 2: Delivering material to the white dwarf's Roche sphere" of the previous chapter "The evolution and delivery of rocky extra-solar materials to white dwarfs". Debris disks, transits from disintegrating bodies, and intact planets have all been detected around white dwarfs (e.g., Jura et al. 2007; Vanderburg et al. 2015, 2020). Perhaps the best example supporting the asteroid tidal disruption theory is the white dwarf WD 1145+017, which has a heavily polluted atmosphere, a circumstellar disk made from dust and gas, and variable transit features from a disintegrating asteroid (Vanderburg et al. 2015; Xu et al. 2016; Rappaport et al. 2016). Since then, many more systems similar to WD 1145+017 have been detected (e.g., Vanderbosch et al. 2020). In addition, Cunningham et al. (2019) reported the first X-ray detection of the polluted white dwarf G29–38, which directly confirms that the white dwarf is actively accreting.

The previous chapter presents the evolution, dynamics, and sizes of these pollutant bodies and describes how this material ultimately ends up in the atmosphere of these white dwarfs. This chapter will describe how the chemical autopsies are conducted, and what is learnt about exoplanetary material from polluted white dwarfs.

## 2. METHODS

In this section, we describe observations of polluted white dwarfs, white dwarf modelling, and how to derive the composition of the planetary bodies from the spectrum of a polluted white dwarf.

---

[2] In this chapter, "heavy elements" refer to all elements heavier than helium.

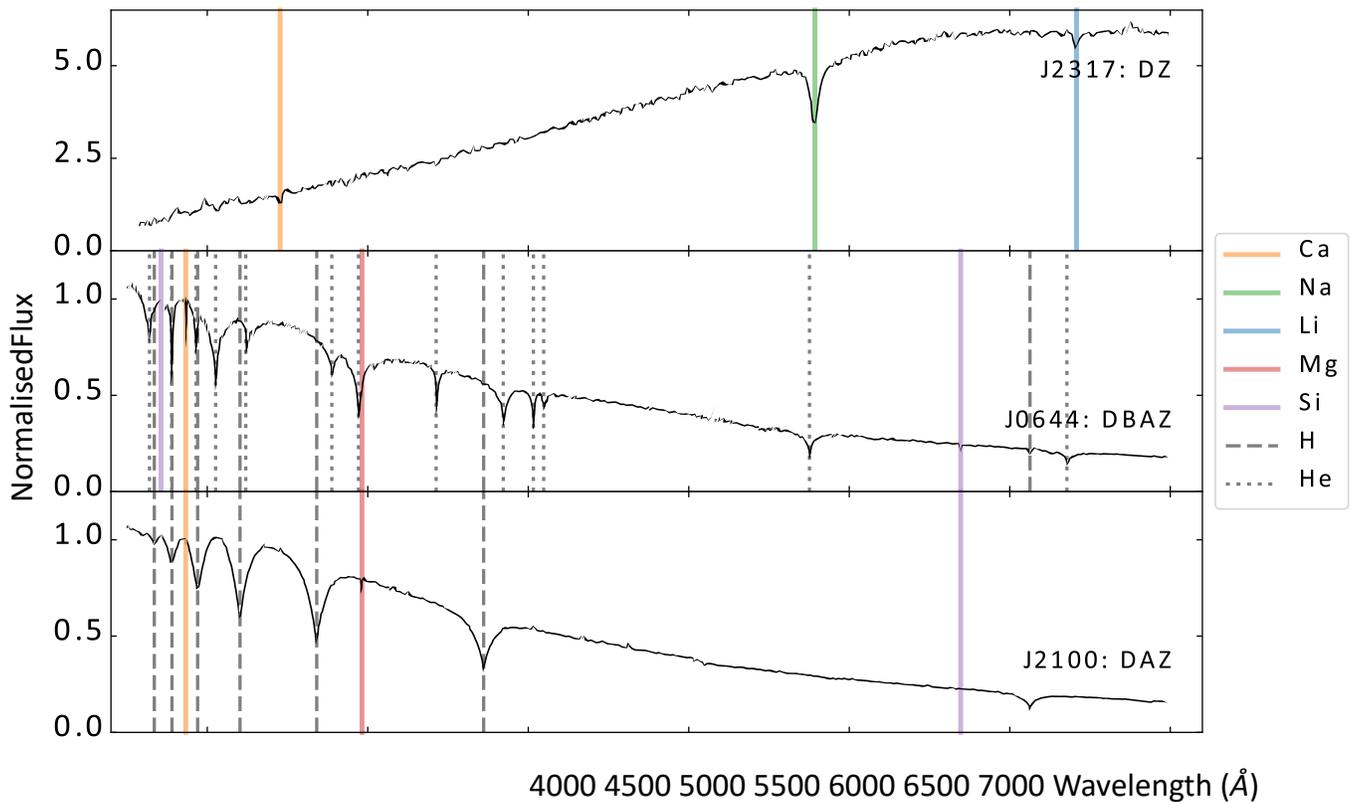

**Figure 2.** Normalized flux versus wavelength for three polluted white dwarfs with different spectral types: WDJ231726.74+183052.75 ($T_{eff}$ = 4600K), a cool DZ white dwarf, GaiaJ0644−0352 ($T_{eff}$ = 18350K), a hotter DBAZ white dwarf, and GaiaJ2100+2122 ($T_{eff}$ = 25570K) a hot DAZ white dwarf. The dashed lines mark observed absorption lines of hydrogen, the dotted lines mark observed absorption lines of helium, and the colored lines highlight observed lines of heavy elements from Ca, Na, Li, Mg and Si. The spectrum of WDJ231726.74+183052.75 is from the OSIRIS spectrograph on the Gran Telescopio Canarias and has a spectral resolution R∼1000 (Tremblay et al. 2020; Hollands et al. 2021), and the spectra of GaiaJ0644−0352 and GaiaJ2100+2122 are from the X-shooter instrument on the Very Large Telescope which has R∼5400 on the blue side (< 5600 °A) and R∼8900 on the red side (> 5600 °A)(Rogers et al. 2024). All spectra are median filtered with a box size of five for clarity.

## 2.1. *Observation*

The spectral classification of white dwarfs is based on their optical spectra. It begins with a 'D' to highlight that a white dwarf is a degenerate object, which is supported by electron degeneracy pressure. A fun fact: the more massive a white dwarf is, the smaller it is. The main spectral classifications of white dwarfs that are most relevant to this chapter are: DA, DB, DC, and DZ. Around 80% of white dwarfs are DAs, where strong and broad hydrogen absorption lines are observed in the spectra (as illustrated in Figure 2). DB white dwarfs show strong helium absorption lines, having lost most of their hydrogen due to a late helium-shell flash (also known as as "born-again" episode) during the late phases of stellar evolution (Werner & Herwig 2006) or due to interactions with a stellar companion (Reindl et al. 2014). DC (C stands for "continuum") white dwarfs display no hydrogen or helium lines in their optical spectra; this occurs when a DA white dwarf cools below 5,000K, or when a DB white dwarf cools below 11,000K (Saumon et al. 2022). Therefore, DC white dwarfs could have either a hydrogen or helium dominated atmosphere. The spectra of DZ white dwarfs contain atomic absorption features from heavy elements (typically Ca)[3]. Many white dwarfs have multiple classifications, and for those the letters are listed in order of which spectral features dominate. For example, DAZ and DBZ are DA and DB white dwarfs that also contain absorption features from heavy elements. DAZ, DBZ, and DZ white dwarfs are called polluted white dwarfs. Figure2 shows example spectra from each of these classes of polluted white dwarfs.

Spectral classification is a useful way to classify white dwarfs, but the strongest optical features are not necessarily indicative of the dominant species. All DZs fall in this category – the strongest optical features come from heavy elements, yet the white dwarf has either a hydrogen or helium dominated atmosphere. Another more subtle example is the heavily polluted white dwarf GD 362, whose spectral type is DAZB, even though it has a helium dominated atmosphere. At the 4temperature of GD 362 (10,540 K, Xu et al. 2013), the He lines become very weak and the H lines are stronger than He, even though the H abundance is lower than He.

---

[3] Another common spectral type for cool white dwarfs (< 10,000 K) is DQ, which shows either carbon lines or molecular $C_2$ Swan bands. The carbon is believed to come from convective dredge up of material from the star's interior, rather than external accretion (Dufour et al. 2005; B´edard et al. 2022).

To obtain the abundances of planetary material polluting white dwarfs, it is crucial to obtain high quality spectra. There is a balance between wavelength range, spectral resolution, and signal-to-noise ratio (SNR). Table 1 shows the wavelengths of the strongest spectral lines for each element that have been detected in polluted white dwarfs. The detectability of an element depends on its abundance, the intrinsic strength of the line, the white dwarf parameters, and the characteristics of the instrument. The resolving power of a spectrograph, $R$, is defined by its ability to distinguish between two wavelengths with a difference of $\delta\lambda$, $R = \lambda/\delta\lambda$. The SNR is the ratio of the flux level to the background noise and is used to define the quality of spectra; the longer the exposure time, the larger the SNR. The SNR is roughly proportional to the square root of the exposure time. Therefore, it requires a lot of observing time to significantly improve the SNR of a spectrum. Low-resolution ($R < 5,000$) optical data is faster to obtain for a large sample of white dwarfs due to the shorter exposure times needed and the vast numbers of telescopes and instruments available; however, this will most often reveal just the strongest absorption lines (e.g., calcium H and K lines at 3933.7 and 3968.5 Å). In order to study the composition of planetary material, abundances from multiple elements are required. This generally needs higher resolution data ($R>20,000$) with a good SNR that spans a larger wavelength range (Table 1), which is more difficult to obtain for a large sample of white dwarfs.

Due to the strong hydrogen opacities for DA white dwarfs, absorption lines from heavy elements tend to be found in abundance in DB white dwarfs. For a given abundance, the same spectral lines tend to be stronger in DBZs than DAZs. In addition, polluted white dwarfs where more than 5 elements are discovered tend to be DBZs. Table 1 shows the wavelengths of dominant spectral lines for elements in the optical (here defined as $\lambda > 3000$ Å) versus the ultraviolet ($\lambda < 3000$ Å). For hotter white dwarfs ($T > 10,000$K) where there is sufficient flux from the white dwarf at the ultraviolet wavelengths, spectral lines in the ultraviolet tend to be stronger and more abundant. However, ultraviolet radiation is absorbed from Earth's atmosphere, and is therefore only observable from space. The *Cosmic Origins Spectrograph* on the *Hubble Space Telescope* is the only instrument available right now with enough sensitivity and spectral resolution to study polluted white dwarfs in the ultraviolet. Additionally, the strongest lines for the volatile elements (C, N, O) are all at ultraviolet wavelengths, making these ultraviolet observations crucial for questions regarding volatile loss and habitability in exoplanetary systems.

There is a total of ≈ 1300 polluted white dwarfs with at least one element measured, according to the Montreal White Dwarf Database (Dufour et al. 2017). Most of these were first identified from the Sloan Digital Sky Survey (SDSS), which is a large multi-object spectroscopic survey that covers 3650 – 10400 Å at a spectral resolution around 2000 (e.g., Koester & Kepler 2015). Calcium is the most easily detected element in the optical, and the majority of the polluted white dwarfs only show absorption features from calcium, as shown in Table 1. About 500 polluted white dwarfs have at least two elements detected (typically Ca and Mg). GD 362 is among the most heavily polluted white dwarf and it holds the record of having 17 heavy elements detected in its atmosphere, including C, Na, Mg, Al, Si, S, Ca, Sc, Ti, V, Cr, Mn, Fe, Co, Ni, Cu, and Sr (Xu et al. 2014). It is also worth noting that Be, Li and K have only been recently detected in polluted white dwarfs (Hollands et al. 2021; Kaiser et al. 2021; Klein et al. 2021).

### 2.2. Modeling

Model atmospheres are used to determine elemental abundances in the atmospheres of polluted white dwarfs. Typically, a multi-dimensional grid of models with different effective temperatures $T_{\rm eff}$, surface gravities $\log g$ ($g = GM_*/R_*^2$), and compositions $n_i$ are calculated. The predicted stellar spectra for each model in this grid are then compared to the observed white dwarf spectrum, and the parameters of the star are assumed to correspond to those of the best-fit model. This procedure of using model atmospheres to deduce elemental abundances is not unique to white dwarfs but is also similar to methods employed to derive stellar abundances in general. In this section, we briefly explain how these models are generated.

**Table 1.** Heavy elements that have been detected in polluted white dwarfs, listing the number of white dwarfs with such detections and the dominant spectral lines in the optical (Op) and ultraviolet (UV) for different ionization states of that element. This list collates the most commonly observed lines for each element, with information collected from: Klein et al. (2010, 2011); Gänsicke et al. (2012); Jura et al. (2012); Hoskin et al. (2020); Klein et al. (2021); Kaiser et al. (2021); Hollands et al. (2021); Izquierdo et al. (2021); Johnson et al. (2022); Rogers et al. (2024). Following the convention, the wavelength of the optical and ultraviolet lines are in air and vacuum, respectively. This list is not exhaustive, but can serve as a starting point when identifying key spectral lines for each element.

| Element | No. | Op/UV | Wavelength (Å) |
|---|---|---|---|
| Li | 6 | Op | Li I: 6707.8, 6707.9 |
| Be | 2 | Op | Be II: 3130.4, 3131.1 |
| C | 41 | Op | C II: 4267.3, 6578.0 |
| | | UV | C I: 1140.4, 1261.6, 1329.1, 1329.6 |
| | | | C II: 1334.5, 1335.7 |
| | | | C III: 1174.9, 1175.3, 1175.7, 1176.0, 1176.4 |
| N | 3 | UV | N I: 1243.2 |
| | | | N II: 1084.6, 1085.7 |
| O | 38 | Op | O I: 7771.9, 7774.2, 7775.4, 8446.4 |
| | | UV | O I: 1152.2, 1302.2, 1304.9, 1306.0 |
| Na | 118 | Op | Na I: 5890.0, 5895.9 |
| Mg | 306 | Op | Mg I: 3832.3, 3838.3, 5172.7, 5183.6 |
| | | Op | Mg II: 4481.1, 4481.3, 7877.1, 7896.0 |
| | | UV | Mg II: 1239.9, 1240.9, 2795.5, 2802.7 |
| Al | 30 | Op | Al I: 3944.0, 3961.5 |
| | | Op | Al II: 3586.6, 3587.1, 3587.4, 4663.1 |
| | | UV | Al II: 1191.8, 1725.0 |
| | | UV | Al III: 1379.7, 1384.1, 1854.7, 1862.8 |
| Si | 72 | Op | Si I: 3906.6 |
| | | Op | Si II: 3856.0, 4128.1, 4130.9, 5056.0, 6347.1, 6371.4 |
| | | UV | Si II: 1190.4, 1193.3, 1194.5, 1197.4, 1260.4, 1264.7 |
| | | UV | Si III: 1141.6, 1161.6, 1296.7, 1298.9 |
| | | UV | Si IV: 1393.8, 1402.8 |
| P | 12 | UV | P II: 1154.0, 1159.1, 1249.8 |
| | | UV | P III: 1334.8, 1344.3 |
| S | 17 | UV | S I: 1316.5, 1425.0, 1425.2 |
| | | UV | S II: 1204.3, 1253.8, 1259.5 |
| | | UV | S III: 1194.0, 1194.4 |
| K | 5 | Op | K I: 7664.9, 7699.0 |
| Ca | 1291 | Op | Ca I: 4226.7 |
| | | Op | Ca II: 3933.7, 3968.5, 8498.0, 8542.1, 8662.1 |
| | | UV | Ca II: 1169.0, 1169.2, 1341.9, 1432.5, 1433.8 |
| Sc | 5 | Op | Sc II: 3572.5, 3613.8, 3630.7 |
| | | UV | Sc II: 1418.8 |
| Ti | 59 | Op | Ti II: 3234.5, 3236.6, 3349.0, 3349.4, 3361.2 |
| V | 5 | Op | V II: 3125.3, 3267.7, 3271.1, 3276.1 |
| | | UV | V III: 1148.5, 1149.9 |
| Cr | 91 | Op | Cr II: 3120.4, 3125.0, 3128.7, 3132.1, 3368.0 |
| | | UV | Cr II: 1435.0, 1435.2 |
| | | UV | Cr III: 1136.7, 1146.3, 1247.8, 1252.6, 1259.0, 1261.9, 1263.6 |
| Mn | 17 | Op | Mn II: 3442.0, 3460.3, 3474.0, 3474.1, 3482.9 |
| | | UV | Mn II: 1162.0, 1188.5, 1192.3, 1192.3, 1197.2, 1199.4, 1201.1, 1234.0, 1254.4 |
| | | UV | Mn III: 1174.8, 1177.6, 1179.9, 1183.3, 1183.9 |
| Fe | 286 | Op | Fe I: 3570.1, 3581.2, 3719.9, 3734.9, 3749.5, 3820.4, 3859.9 |
| | | Op | Fe II: 3227.7, 5018.4, 5169.03 |
| | | UV | Fe II: 1143.2, 1144.9, 1358.9 |
| Co | 2 | Op | Co II: 3754.7 |
| Ni | 30 | Op | Ni I: 3524.5, 3619.4 |
| | | Op | Ni II: 3514.0 |
| | | UV | Ni II: 1317.2, 1335.2, 1370.1, 1381.3, 1411.1 |
| Cu | 1 | Op | Cu I: 3247.5, 3274.0 |
| Sr | 3 | Op | Sr II: 4215.5 |

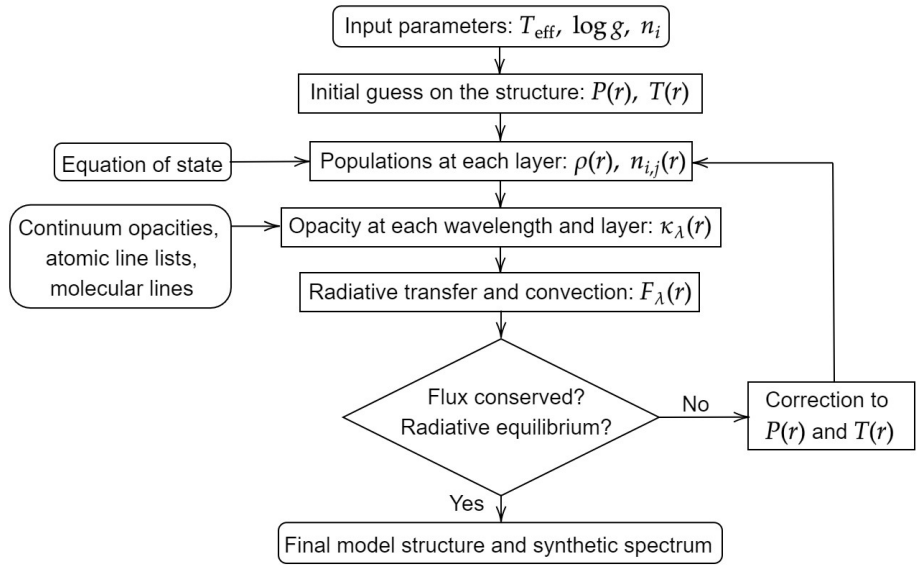

**Figure 3.** Schematic representation of a white dwarf model atmosphere calculation.

Model atmospheres rely on a number of standard approximations. First, the horizontal directions are normally ignored: the atmosphere is treated in a 1-D framework and vertically divided into O($10^2$) layers. These layers are arranged in a way that allows one to capture all the absorption sources that contribute to the star's emerging flux. More specifically, the deepest layer is positioned well below the region where absorption lines are formed, at an optical depth > $10^2$ (i.e., deep enough that photons are absorbed or scattered many times before escaping the atmosphere). Similarly, the uppermost layer is placed high up in the atmosphere (optical depth < $10^{-6}$), above the line-forming region (roughly located between an optical depth of 1 and $10^{-5}$). Because the atmosphere is very thin (∼ $10^2$ m) compared to the star's radius (∼ $10^7$ m), a plane-parallel geometry is assumed.

The structure of the atmosphere is given by the hydrostatic equilibrium equation,

$$\frac{dp}{dr} = -\rho(r)g, \tag{1}$$

where $p(r)$ is the pressure structure of the atmosphere, $\rho(r)$ its density stratification, and $g$ the surface gravity.[4] As Equation (1) does not prescribe the temperature structure $T(r)$, the first step in the calculation of a model atmosphere consists of guessing $T(r)$, which can be achieved using an approximate Hopf function (Mihalas 1978). Equipped with $P(r)$ and $T(r)$, an equation of state can be solved at each layer to obtain the abundances of all species (free electrons, ions, molecules). Note that the atmosphere is normally assumed to be chemically homogeneous, meaning that each layer has the same elemental composition (e.g., the ratio of Ca to He nuclei is constant). This elemental composition is an input parameter of the model atmosphere calculation. However, since $P$ and $T$ change throughout the atmosphere, the ionic and molecular populations vary (e.g., the ratio of singly ionized to neutral Ca changes). For most white dwarf atmospheres, the equation of state physics is well known (the ideal gas law and the Saha ionization equation are generally applicable)[5], but we will discuss the important exception of old, cool white dwarfs in Section "Constitutive physics uncertainties".

Once the abundances of each species is determined, the next step consists of calculating the radiative opacity of the mixture. Many contributions must be considered: Thomson scattering from electrons, Rayleigh scattering from atoms and molecules, bound–free absorption, free–free absorption, bound–bound absorption (spectral lines), molecular bands, and collision-induced absorption. In the current context, spectral absorption lines (and to a lesser extent, molecular bands) represent the most interesting aspect of the opacity calculation, as they are the features that are directly used to infer elemental compositions. Accordingly, we will focus on this aspect of the opacity calculation in what follows.

That said, we note that in general it is the other opacity sources (which do not produce salient spectral features but instead a more continuous absorption) that dominate the total opacity of the mixture and therefore control the thermodynamic structure of the atmosphere.

Absorption lines are included using compilations such as the Kurucz lines list[6], the Vienna Atomic Line Database (VALD)[7], or the NIST atomic spectra database[8]. These sources provide the strengths and wavelengths of all absorption lines observed in white dwarfs, although the uncertainties on the line lists can be quite large. For each line, an absorption profile that accounts for both temperature (Doppler) and collisional broadening is calculated. We will see in Section "Constitutive physics uncertainties" that this task is still a challenge for very cool white dwarfs.

Once the opacity of each atmospheric layer is determined, the theory of radiative transfer enables the calculation of the radiative flux at each level of the vertical stratification. Since $T(r)$ was initially guessed, the energy flux[9] is generally not conserved throughout the atmosphere and radiative equilibrium is not attained. The temperature structure of the atmosphere must therefore be corrected until each layer transports a flux corresponding to $\sigma T_{\text{eff}}^4$, where $\sigma$ is the Stefan–Boltzmann constant (see Bergeron et al. 1991 for details on the numerical implementation of this correction procedure). The steps described above are repeated until a physical solution is reached (see Figure 3). This final model structure is then used to generate a high-resolution synthetic spectrum that can be compared to observations after convolving it to the instrument's response function.

Once a multi-dimensional grid of model spectra has been calculated, how does one measure the elemental abundances of a given star? An iterative procedure that alternates between fitting $T_{\text{eff}}/\log g$ and the individual abundances is often employed (e.g., Dufour et al. 2007; Coutu et al. 2019; Blouin 2020). $T_{\text{eff}}$ and $\log g$ can first be estimated using photometric data (measurements of the object's intensity across specific bandpasses) and a parallax measurement (which gives the distance $D$ separating the star from the Sun). With this approach, now widely used thanks to the precise parallaxes provided by the Gaia mission (Gaia Collaboration 2016), the solid angle $\pi(R_*/D)^2$ and $T_{\text{eff}}$ are directly adjusted to the photometric data points using a $\chi^2$ minimization algorithm. Given the known white dwarf mass-radius relationship, this also yields the surface gravity $\log g$. Once a photometric solution is found, the individual abundances are adjusted to fit the absorption lines detected in the spectroscopic data (Figure 4). After that, the photometric procedure is repeated once more, as the addition of metals to the model atmosphere generally changes the overall shape of the emerging spectrum in a way that requires a different $T_{\text{eff}}/\log g$ to fit the photometry. Both steps (fitting the photometry and the spectroscopy) are repeated until internal consistency is reached.

---

[4] In practice, Equation 1 and its different profiles are parameterized as a function of the optical depth $\tau$ instead of the geometrical depth $r$. We use $r$ here for simplicity.

[5] Local thermodynamic equilibrium is normally assumed, which is almost always well justified.

[6] http://kurucz.harvard.edu/linelists.html

[7] http://vald.astro.uu.se/

[8] https://physics.nist.gov/PhysRefData/ASD/lines form.html

[9] For most polluted white dwarfs (with the notable exception of those with hydrogen-dominated atmospheres and $T_{\text{eff}} \gtrsim 18{,}000\,\text{K}$), the energy flux in the atmosphere also includes a convective component.

2.3. *Composition of the Accreting Material*

Once we have the white dwarf atmospheric abundances, we can calculate the mass of the polluting material in the white dwarf's atmosphere. This mass is a lower limit because it only represents the amount of material that is currently in the white dwarf's atmosphere. For cooler DB white dwarfs, due to the transparent atmospheres, more mass is maintained in the convection zone, and this can provide more accurate estimates of the total mass accreted, albeit this is still a lower limit. The mass of heavy elements in the convection zone assuming a bulk Earth like composition is approximately $10^{20}$ – $10^{25}$ g (Farihi et al. 2010; Girven et al. 2012) Harrison et al. (2021a) analyzed thirteen heavily polluted white dwarfs and estimated the parent body mass ranges from $10^{23}$ – $7 \times 10^{25}$ g, as shown in Figure 5).

In addition, we can infer the elemental composition of the planetary body. It is often assumed that the white dwarf atmosphere is dominated by one single parent body. The accretion event in the white dwarf's atmosphere is a dynamic process, with material continuously falling onto the white dwarf and settling out of the upper atmosphere that is visible to us, as illustrated in Figure 1.[10] For one parent body, the accretion stage can be divided into three distinctive phases, i.e., the build-up state, the steady state, and the declining state (Koester 2009). Figure 6 illustrates the possible accretion history for a polluted white dwarf. In the build up state, the observed atmospheric abundance is the same as the composition of the parent body. No additional correction is needed for this stage. In the steady state, the observed composition is modified by the diffusion time of each element. This typically changes the relative abundance ratios of the elements by factors of a few. While in the declining state, the observed composition can be quite different from the original composition and the abundance ratios can differ by up to several orders of magnitude.

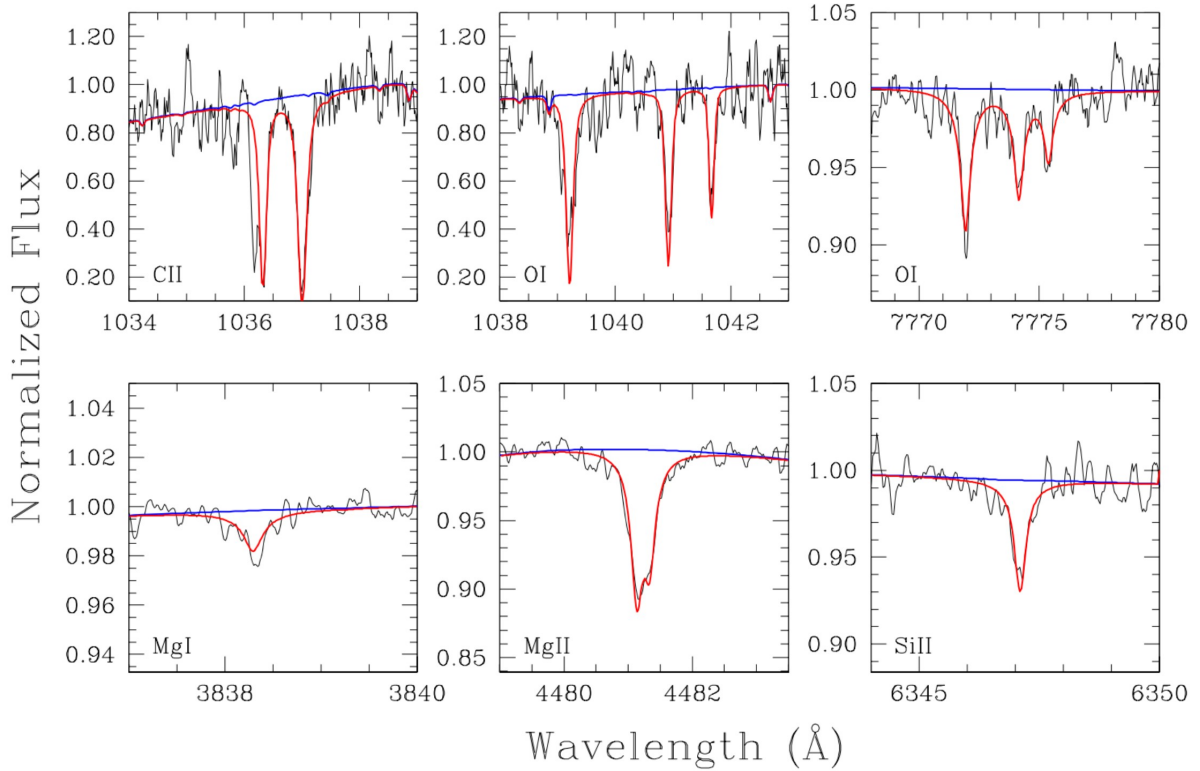

**Figure 4.** Illustration of the spectroscopic fitting procedure that yields the elemental abundances. The red line is the best-fit model to the observed polluted white dwarf spectrum (in black), and the blue line is the same model but where the abundance of the element indicated on each panel has been forced to zero. The UV data (including C II and O I) are from the FUSE satellite and the optical data are from Keck/HIRES. Figure taken from Klein et al. 2021 (© AAS, reproduced with permission).

---

[10] In very hot white dwarfs ($T_{\rm eff} \gtrsim 25{,}000$K), radiative levitation can complicate this picture, as the upward force exerted by radiation on certain ions can cause them to remain at specific atmospheric depths where the radiative force counterbalances gravitational settling.

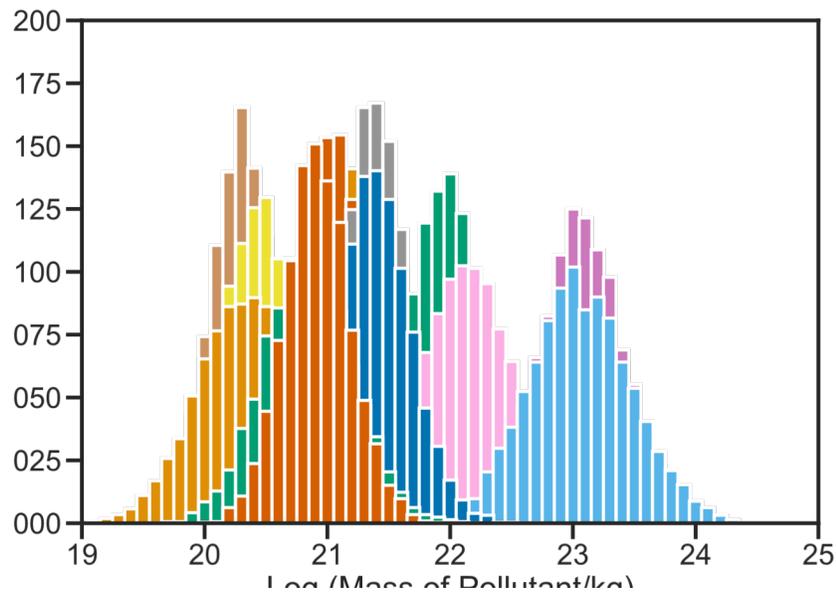

**Figure 5.** The posterior distribution of the total mass accreted onto thirteen polluted white dwarfs from Harrison et al. (2021a). The total mass varies from half of Vesta to the Moon.

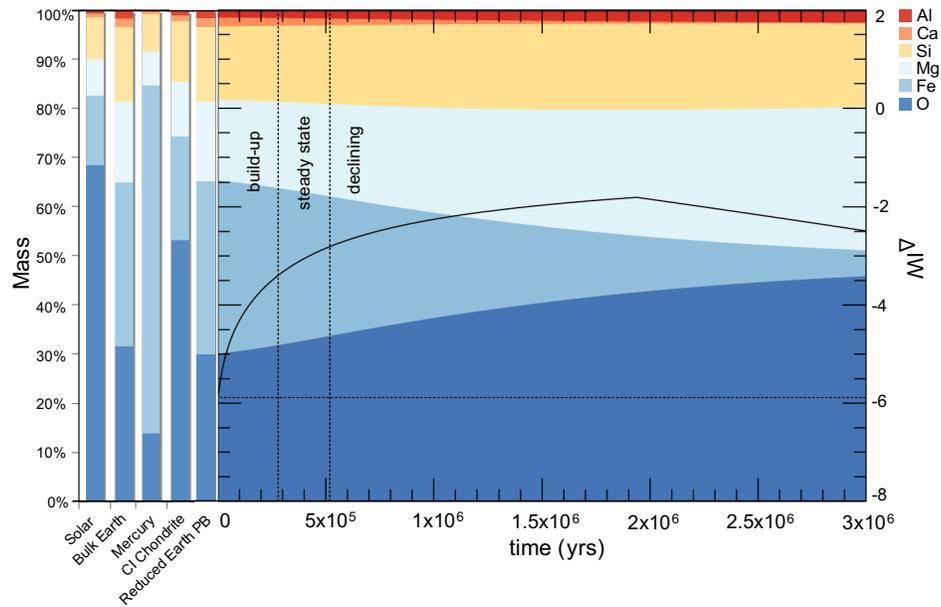

**Figure 6.** Possible accretion history for a polluted white dwarf from assuming it is accreting from one single object. The parent body (PB) has a reduced Earth composition. Depending on the accretion state (i.e., build-up, steady state, declining), the observed composition in the polluted white dwarfs is different. Therefore, additional corrections may be needed to infer the 'real' composition of the disintegrating parent body. Figure taken from Doyle et al. 2020 (© AAS, reproduced with permission).

For example, if the white dwarf is observed at the time $2.5 \times 10^6$ yr, one may conclude that the accreting material is O-rich and Fe-poor. In reality, this is just the result of differential settling; oxygen has the longest diffusion time and is still mostly in the atmosphere, while iron has largely diffused out of the atmosphere. Generally speaking, if a white dwarf has an infrared excess from a dust disk, we can assume that the system is either in the build up state or the steady state. If the system of interest is a hot DA white dwarf with very short settling times ($\lesssim 1$ year), we often assume that it is in a steady state. In these cases, it is straightforward to derive the composition of the parent body. It is a lot more difficult to derive the planetary composition if the system is in a declining phase. One way around this is to model all the potential compositions of the planetary body by sampling all the accretion scenarios, as has been done in, e.g., Swan et al. (2019); Doyle et al. (2020); Buchan et al. (2022). The best practice would be to try to explain the observed composition with the simplest scenario, before invoking unique and special explanations.

It is a lot harder to derive the 'real' composition of the accreting material if a white dwarf is accreting from multiple planetary bodies at the same time. To make progress on the theoretical front, the accretion theory needs a coupled treatment of dust and gas, which is still an area of active research (Okuya et al. 2023). Strong variability appears to be common in the circumstellar dust and gas around white dwarfs (Swan et al. 2019; Dennihy et al. 2020). By analyzing a large unbiased sample of polluted white dwarfs, Wyatt et al. (2014) found that the accretion is continuous (rather than stochastic) for small planetary bodies ($6.6 \times 10^{19}$ g, or 35 km diameter for 3 g cm$^{-3}$), and this may be the dominant source of pollution. On the other hand, no variability in the accretion rate has been confidently detected, corroborating the picture that accretion appears to be in a steady state (Debes & Kilic 2010; Johnson et al. 2022). There could also be chemical alterations to the composition of the planetary body prior to the accretion onto the white dwarf, as discussed in the sections below and the previous chapter "Evolution and Delivery of Rocky Extra-Solar Materials to White Dwarfs".

## 3. INTERPRETATION

In this section, we describe commonly used methods to interpret the abundances of the extrasolar planetary material that has been accreted onto white dwarfs. We begin by discussing some of the key chemical, thermal, and physical processes that determine the composition of the planet, and then discuss how these processes are investigated using polluted white dwarfs.

### 3.1. *Key Processes*

Planets are expected to inherit their composition from the interstellar cloud under which the star and planet formed. The composition of this planetary material may subsequently be altered by chemical, thermal, and physical processes, such as: heating during formation, post-nebula processing, differentiation, and collisional fragmentation. If no further processing occurs then the planetary material accreted would be 'primitive'; in the solar system, CI chondrites are most similar to the composition of the sun and have gone through little subsequent processing.

**Heating during formation:** The initial volatile budget of the planetary material is not set solely from the interstellar cloud, but additionally by its formation location with respect to the various snow lines, this leads to incomplete condensation of some elements (e.g. Pontoppidan et al. 2014). Planetary embryos can additionally accrete from a range of radial locations, called a 'feeding zone'. In the solar system, the volatile inventories of the inner terrestrial planets versus the outer gas and ice giants are broadly consistent with the positions of the snow lines.

**Volatile loss:** Post-nebula volatile loss occurs when a planet experiences intense heating after the protoplanetary disk has evaporated, for example, due to energetic collisions (Safronov & Zvjagina 1969), or heat released from the decay of short-lived radioisotopes (e.g. Urey 1955). This leads to a partial or full magma ocean phase in which volatiles can be degassed (e.g. Elkins-Tanton 2012). The volatile inventory from incomplete condensation in the protoplanetary disk and post-nebula volatilization are governed by different pressure, temperature, and oxidisation conditions, and therefore, these volatile depletion processes imprint uniquely on the abundance pattern of the planetary body (Harrison et al. 2021c).

**Core-mantle differentiation:** If there is sufficient heating that large-scale melting of the parent body occurs, the segregation of the iron melt leads to the formation of a core and a mantle which form under the influence of the internal pressure and oxygen fugacity (e.g. Trønnes et al. 2019). Siderophilic elements migrate to the core, and lithophilic elements tend to the surface of the body, depending on the exact conditions under which this occurs bulk compositional differences can arise. Subsequent collisions, or other processing that leads to fragmentation, can change the core to mantle ratio in a planetesimal.

**Collisional fragmentation:** Collisions during the formation and evolution of our solar system are fundamental to the design and composition of the planetary bodies. For example, the proto-Earth collided with a proto-planet approximately the size of Mars which resulted in the Earth-moon system we see today (Canup & Asphaug 2001). Collisions and fragmentation can change the relative amounts of different species, especially if the body has a compositional structure such as it being differentiated. Certain classes of meteorites (e.g. iron meteorites) are the collisional fragments of larger differentiated planetesimals.

### 3.2. *Comparison with Meteorites*

When it became clear that polluted white dwarfs are accreting from extrasolar planetary material, the natural step was to compare the measured composition with those of rocky bodies in the solar system. The best composition database in the solar system comes from meteorites, which are fragments of minor bodies. When enough elements are detected in a polluted white dwarf, it is possible to do a direct comparison with the meteorites, look for the best match, and infer the formation scenario of the parent body. For example, GD 40 is the first polluted white dwarf for which all the major rock forming elements (i.e. Ca, Mg, Fe, Si, and O) were detected. Klein et al. (2010) found that the overall constitution of the accreting material in GD 40 is similar to bulk Earth, but the Mg/Si ratio is smaller than the value for bulk Earth and nearby stars, which may be due to accretion from a differentiated parent body, as discussed in the previous section. Using a $\chi^2$ comparison, Xu et al. (2013) found that the best solar system analog to the material accreting onto GD 362 is mesosiderite, which is a rare type of stony-iron meteorite and is a mixture of crust and core material. G 238–44 has 10 heavy elements detected and the observed composition has no counterpart in the solar system. The best match is a mixture of iron-rich Mercury like material and an analogy of a Kuiper Belt Object (Johnson et al. 2022). Swan et al. (2019, 2023) developed methods that compare the abundances of polluted white dwarfs to solar system bodies to find the most likely compositional match, with one showing that the best match is the rare achondrite, acapulcoite.

Element ratio plots are the most widely used method to compare polluted white dwarfs with other objects. In the very early days, Jura (2006) compared the carbon-to-iron ratio between three polluted white dwarfs, the Sun, comet Halley, Earth's crust, and different meteorites. They found that carbon is depleted by more than a factor of 10 in polluted white dwarfs compared to the solar value, rejecting the interstellar accretion theory and providing a strong support for the asteroid accretion theory. Here, we revisit the commonly used element ratio figures, since the number of polluted white dwarfs has increased significantly since the last major review paper by Jura & Young (2014). The abundances of polluted white dwarfs are assembled from the literature (Koester & Wolff 2000; Zuckerman et al. 2007; Klein et al. 2011; Melis et al. 2011; Zuckerman et al. 2011; Dufour et al. 2012; Gänsicke et al. 2012; Jura et al. 2012; Kawka & Vennes 2012; Melis et al. 2012; Farihi et al. 2013; Xu et al. 2013; Vennes & Kawka 2013; Xu et al. 2014; Jura et al. 2015; Raddi et al. 2015; Farihi et al. 2016; Kawka & Vennes 2016; Gentile Fusillo et al. 2017; Hollands et al. 2017; Melis & Dufour 2017; Xu et al. 2017; Blouin et al. 2018; Swan et al. 2019; Xu et al. 2019; Fortin-Archambault et al. 2020; Hoskin et al. 2020; Kaiser et al. 2021; González Egea et al. 2021; Klein et al. 2021; Izquierdo et al. 2021; Elms et al. 2022; Hollands et al. 2021; Johnson et al. 2022; Doyle et al. 2023; Izquierdo et al. 2023; Swan et al. 2023; Rogers et al. 2024; Vennes et al. 2024). We assume the polluted white dwarfs are all in the build up phase (see section "Composition of the Accreting Material") and apply no additional correction to the observed abundances. As a comparison, we have assembled a sample of FG main-sequence stars from the Hypatia catalog (Hinkel et al. 2014). The solar values are taken from Lodders et al. (2009). We also include a large number of meteorites from Nittler et al. (2004); Alexander (2019a,b) and they are broadly separated into three categories: (1) "primitive" chondrites, which include carbonaceous chondrites, ordinary chondrites, and enstatite chondrites; (2) primitive achondrites, which include ureilites (URE), brachinites (BRA), acapulcoites and lodranites (ACA-LOD), winonaites and IAB and IICD irons (WIN-IAB); and (3) differentiated achondrites, which include angrites (ANG), aubrites (AUB), howardite-eucrite-diogenite (HED), mesosiderite (MES), palasites (PAL), and IIE iron meteorites (see more discussions of different meteorite groups in the Chapter "Meteorites and Planets Formation"). In addition, we also included the composition of bulk earth (All`egre et al. 2001) and bulk Vesta (Toplis et al. 2013), which is assumed to be the parent body of the HED meteorite. The results are shown in Figures 7 to 10 and now we discuss some general trends among the abundance patterns.

**Volatile elements (C, S)**: As shown in Figure 7, volatile elements are often depleted in polluted white dwarfs compared to the abundances of FG stars, similar to the values of the meteorites. The only system that has accreted solar carbon and sulfur abundances is WD 1425+540, which is also high in N and has been suggested to accrete from an analog of a Kuiper belt object (Xu et al. 2016). PG 1225-079 remains the only object that is strongly depleted in S but only moderately depleted in carbon. The observed composition does not match with any meteorite and possibly it needs a blend of two objects to explain the abundance pattern (Xu et al. 2013). Interestingly, there are two objects with an overabundance of S (i.e. PG 0843+517 and Gaia J0611-6931), and as PG 0843+517 also shows an overabundance of Fe (as shown in Figure 10), it has been suggested that FeS may be a major consistent in the polluting material (Gänsicke et al. 2012).

**Fe and Moderately Siderophile elements (Cr, Mn, Ni)**: Figure 8 shows the spread of the ratios of moderately siderophile elements is pretty small, actually smaller than the observed range in meteorites. This indicates that the core formation process may be similar in both the extrasolar planetary systems and the solar system. Interestingly, WD 1145+017, a white dwarf with an actively disintegrating object in orbit, stands out as having low Mn/Fe and Cr/Fe ratios. The mass fraction of Fe is consistent with CI chondrite, and the depletion of Mn and Cr may be due to their low condensation temperatures (Fortin-Archambault et al. 2020). GD 378 stands out as having a lower Cr/Fe ratio and it is also one of the only two white dwarfs with detection of Be (Klein et al. 2021). In fact, the Be abundance in GD 378 is about two orders of magnitudes higher than the chondritic value and it may be a result of accretion of icy moons (Doyle et al. 2021).

**Lithophile Elements (Al, Ca, Ti, Na)**: Figure 9 shows the abundance ratios of different lithophile elements. For the Ti/Ca and Al/Ca figure, the spread in polluted white dwarfs is smaller than the range in all meteorites. Ti/Ca is much lower than in chondrites and more comparable to other meteorites. In addition, there are a few white dwarfs with enhanced

Na/Ca ratio, which could be a signature of accretion from crust material. However, this is not supported by the abundances of other elements (such as the stringent upper limit on Si, Swan et al. 2019). On the other hand, there are some very cool and old DZs with highly enhanced Li abundances and one scenario is that they have accreted crustal material (Hollands et al. 2021; Elms et al. 2022). However, it is difficult to pinpoint the exact scenario because neither Al nor Si is detected in these objects. The enhanced Li abundance relative to the other elements could be due to the other elements being more scarce when these old systems were formed (Kaiser et al. 2021). The large Na/Ca ratio observed in WD J2356–209 is interpreted as a possible accretion from comet 67P like object (Blouin et al. 2019).

**Differentiation & Collision**: If a white dwarf accretes an entire asteroid that is differentiated, we may not be able to distinguish it from chondritic material. However, if some fragments of a differentiated body are accreted, it would display distinctive chemical signatures – that is what have been observed in polluted white dwarfs. As shown in Figure 10, there is a large spread in the Fe/Si and Fe/Al ratios in polluted white dwarfs, much larger than the spread in FG stars and the spread in Mg/Si and Si/Al ratios. A natural explanation for the large spread of the Fe abundance is accretion from a differentiated parent body, as first proposed in Jura et al. (2013). NLTT 43806 has the lowest Fe/Al ratio of all polluted white dwarfs and it is a good candidate for accretion of crust material (Zuckerman et al. 2011). Different analyses using polluted white dwarfs all found that differentiation and collision appear to be common in extrasolar planetary systems. For example, Bonsor et al. (2020) focused on the Ca/Fe measurements in 179 white dwarfs and found $66^{+4}_{-6}$ % of the sample must have accreted remnants of differentiated bodies to explain the distribution of Ca/Fe. Doyle et al. (2020) focused on 16 white dwarfs where all the major rock forming elements (i.e., Al, Ca, Si, Mg, Fe and O) are detected and found that while most objects were formed under oxidizing conditions, about 25% were consistent with more reduced parent bodies. The prevalence of differentiation in extrasolar planetary systems can be used to provide an independent constraint on the formation timescale of planetesimals (Bonsor et al. 2023).

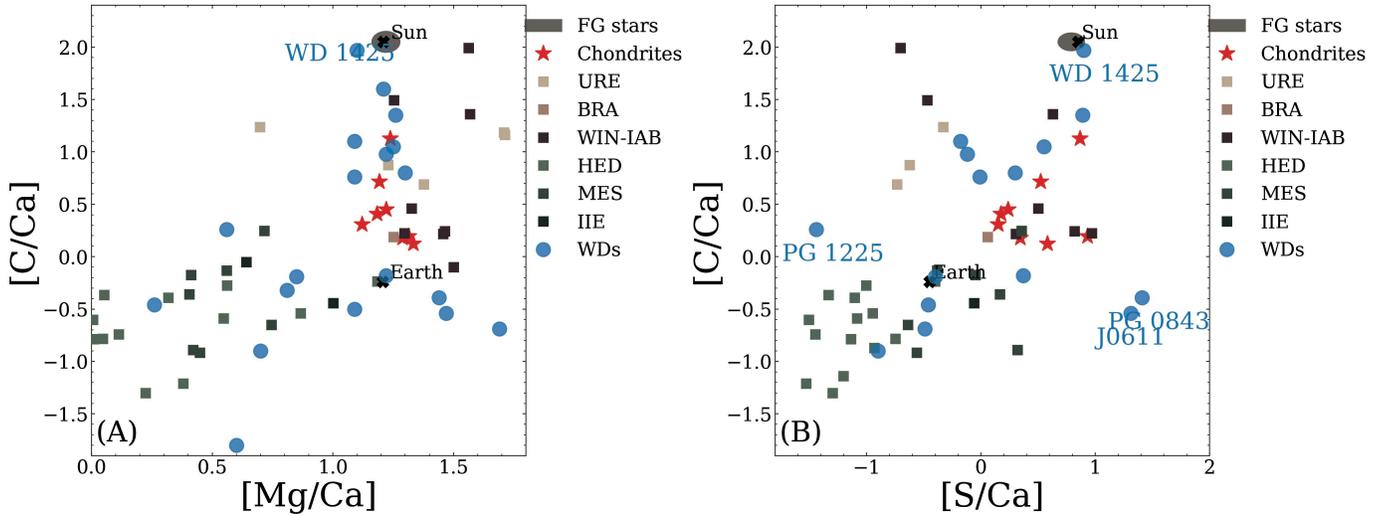

**Figure 7.** Logarithmic number ratios of different elements. The polluted white dwarfs are shown as blue dots and the average uncertainties are about ≈ 0.2 dex in the abundance ratios. The 95 percentile of the abundance ratios of the FG stars are shown as the grey ellipse. The chondrites are shown as red stars while other meteorites are shown as squares. When the abundances are available, the solar value, bulk Earth, and bulk Vesta are also shown as black crosses. Most polluted white dwarfs show a depletion of volatile elements.

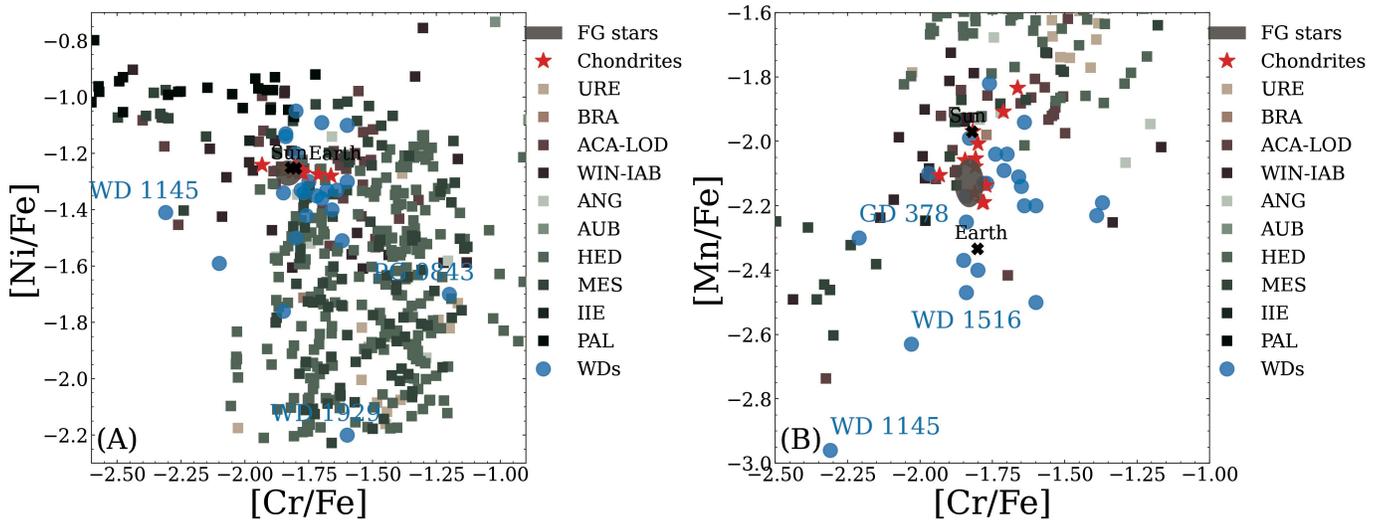

**Figure 8.** Similar as Figure 7 but for moderately siderophile elements. The spread observed in polluted white dwarfs is smaller than the spread of various meteorites.

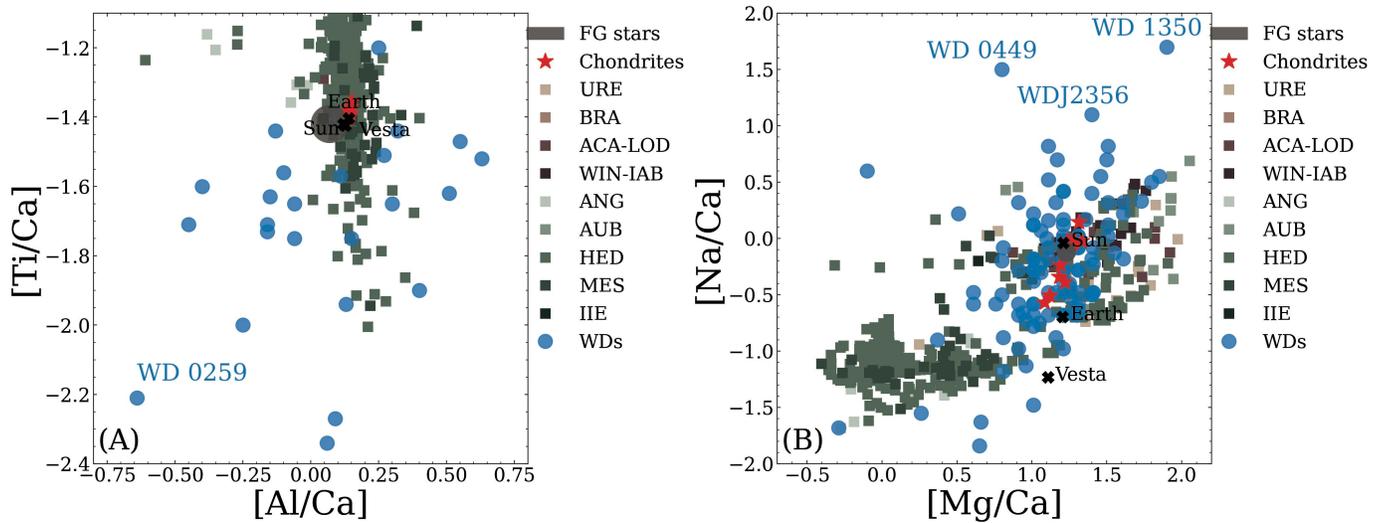

**Figure 9.** Similar as Figure 7 but for lithophile elements. The Ti/Ca ratios in polluted white dwarfs tend to be lower than the chondritic value and FG stars. The high Na/Ca ratios in some polluted white dwarfs cannot be simply interpreted as crust remnants; measurements of additional elements are needed to confirm this scenario.

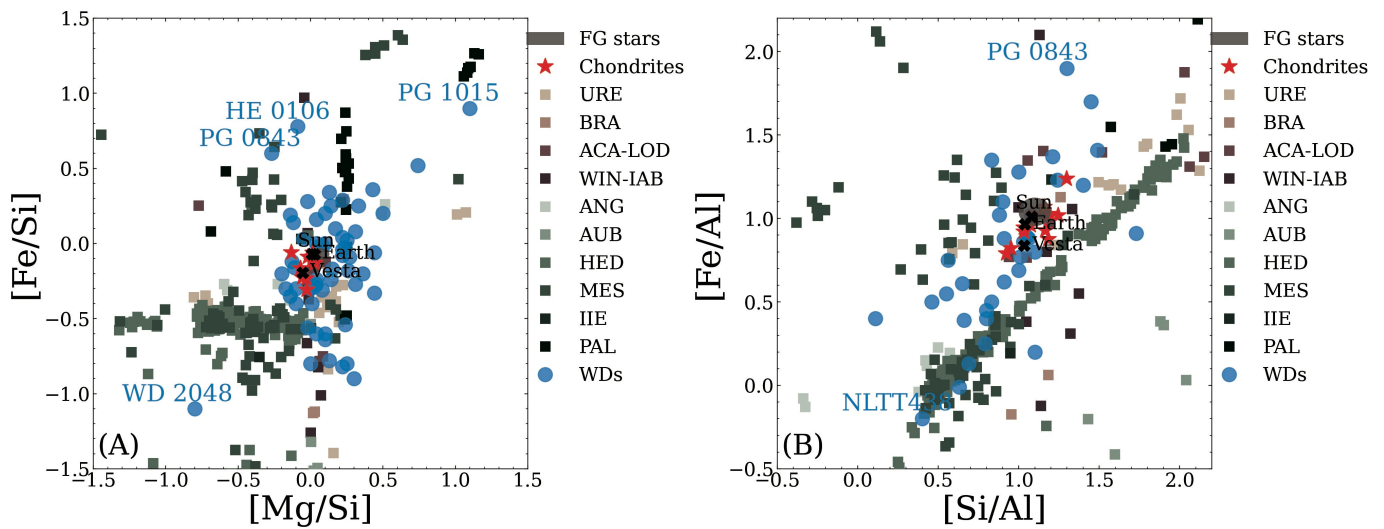

**Figure 10.** Similar as Figure 7 but for a different set of elements. For polluted white dwarfs, while most points cluster around the values for FG stars and chondrites, there are some objects with particularly high Fe and low Fe abundances. A natural explanation is that these white dwarfs have accreted fragments of a differentiated parent body.

In summary, the element ratio plot is very useful in comparing a large number of objects and identifying order-of-magnitude differences. The main drawback is that only a few elements can be displayed at a time, and it is hard to conclude the nature of a given object without looking at all the detected elements. In addition, the figure does not show absolute numbers; so similarties may appear to exist between different groups, though it is actually different. Another issue is that meteorite measurements are typically done on a small amount of material and it is difficult to infer back the bulk composition of the parent body. However, those are still the best composition database available for comparison.

| Model | Free Parameters | Schematic Diagram |
|---|---|---|
| (1) Initial Composition | Stellar metallicity: $[Fe/H]_{index}$<br>Time since accretion started: $t$<br>Accretion event lifetime: $t_{event}$<br>Pollution Fraction, $f_{pol}$ |  |
| (2) Volatile Depletion | Formation distance, $d_{formation}$ | 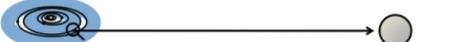 |
| (3) Feeding Zone | Feeding zone size, $z_{formation}$ | 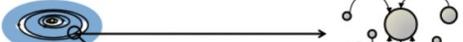 |
| (4) Core-Mantle Differentiation | Fragment core fraction, $f_c$<br>Core-mantle pressure, $P$<br>Core-mantle oxygen fugacity, $f_{O_2}$ | 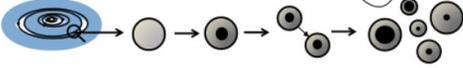 |

**Figure 11.** Table showing the models that are fitted in PyllutedWD. Model 1 is the 'primitive' model that only depends on the initial composition of the stellar nebula and the white dwarf accretion parameters; these parameters are always fitted. Models 2-4 show additional parameters that the Bayesian framework considers to explain the pollutant abundances. Figure adapted from Harrison et al. (2018) with updated parameters from Buchan et al. (2022).

To take this one step further, Putirka & Xu (2021) attempted a mineralogy classification on the material accreted onto polluted white dwarfs. They found no evidence for accretion of continental crust in polluted white dwarfs, and found that some white dwarfs have exotic compositions with no analogs in the solar system. However, Trierweiler et al. (2023) found that the uncertainties on the abundance measurements of polluted white dwarfs may be too big to make a definitive statement. Refer to the next chapter "Exoplanet Mineralogy: Methods & Error Analysis" for a more in-depth discussion on the exoplanet mineralogy analysis.

### 3.3. *Modeling Exoplanetary Abundances and Exogeology*

Given that the history of a planetary body defines its composition and internal structure, considering the body from formation all the way to the accretion onto the white dwarf is crucial to truly understanding the processes affecting the abundance patterns. To move the polluted white dwarf analysis beyond directly comparing with the meteorites and solar system bodies, Harrison et al. (2018, 2021b) pioneered work that took the abundances of the planetary material and traced the history of this body back to its birth environment. Based off this work, Buchan et al. (2022) developed the open source package PyllutedWD[11] which finds the most likely explanation for the observed composition of a planetary body that has polluted a white dwarf, incorporating all possible accretion histories in the white dwarfs' atmosphere with various processes that planetary bodies may experience. The model compares the likelihood that the abundances can be invoked by a basic primitive model, where the abundances assume stellar like material (Brewer & Fischer 2016), with a range of more complex models (listed in Figure 11) incorporating geochemical processes such as those discussed in section "Key Processes". These models allow for improved understanding of how the abundances in a polluted white dwarf atmosphere relate to the parent body formation, collisional, and geological history, as well as provide a framework to statistically study a large population of polluted white dwarfs. Often the overall conclusions about any particular system are consistent with previous analysis if there are solar system analogs. The following discusses key results about these geochemical processes inferred from the modelling.

**Evidence for heating during formation:** Harrison et al. (2018) found that by modelling the accreted abundances of PG 1225-079, the abundances of the refractory versus moderately volatile elements is most consistent with the accretion of a body that was extremely dry and formed in a region of the protoplanetary disk where temperatures reached above 1400K – such an object does not exist in the solar system. It is consistent with the meteorite comparison shown in Figure 7(B). Additionally, by modelling the Hollands et al. (2017) DZ sample of 202 polluted white dwarfs, Harrison et al. (2021b) discovered that 11 white dwarfs were found to show depletion of volatiles, and the best-fitting models found that heating (described in the form of incomplete condensation of volatile species) was required to explain this abundance trend. Three of these systems required further heating such that even more moderately volatile species (e.g., Mg) did not condense.

---
[11] https://github.com/andrewmbuchan4/PyllutedWD Public

**Evidence for post-nebula volatile loss:** GD 362 has the highest number of elements detected in any one polluted white dwarf. To explain the high Mn/Na ratio, Harrison et al. (2021c) showed that the body accreted must have experienced post-nebula volatilization. This is because the volatility of Mn and Na depends on the temperature, pressure, and oxidisation conditions, and so Na becomes more volatile during post-nebula volatilization processes which leads to an increased Mn/Na ratio in the planetary body.

**Evidence for core-mantle differentiation:** As outlined in the previous section "Comparison with Meteorites", by comparison with solar system analogues a number of white dwarfs were inferred to be accreting fragments of coremantle differentiated bodies. By re-analysing the Hollands et al. (2017) sample, Harrison et al. (2021b) inferred that 65 out of 202 polluted white dwarfs studied have a preference for accreting a fragment of a core-mantle differentiated body, showing that differentiation and collisions appear to be commonplace in exoplanetary systems. Additionally, Buchan et al. (2022) found that from a sample of 42 white dwarfs, 14 showed evidence for the accretion of core-mantle differentiated material. The sample was selected to include those polluted white dwarfs that had a Fe detection and a detection/upper bound for one or more of: Cr, Ni, and Si, the importance of which is discussed below.

**Constraining size of the parent body:** Using the 14 polluted white dwarfs in the sample found to be accreting core or mantle rich fragments of larger differentiated parent bodies, Buchan et al. (2022) used the abundance pattern to constrain the sizes of these differentiated parent bodies. The amount of Ni, Cr, and Si that partition into the core or mantle is a function of the oxygen fugacity and pressure at which core-mantle differentiation occurred; therefore, studying the abundance patterns of fragments of differentiated objects that have accreted onto polluted white dwarfs can reveal the size of the parent body (pre-fragmentation). Three systems (WD0449−259, WD1350−162, and WD2105−820) were found to be best explained by the accretion of a core-rich fragment, and the pressure (as constrained in Model (4) in Figure 11) at which the parent bodies underwent core-mantle differentiation was low (Figure 12a). This implies that the parent body was small (i.e. asteroid sized) rather than on a planetary sized scale. Two polluted white dwarfs (GD61 and WD0446−255) were best explained by the accretion of a mantle-rich fragment at which differentiation occurred at high pressures in the parent body (Figure 12b). This suggests that the parent bodies (pre-fragmentation) were larger, with masses 0.61 Earth mass $M_\oplus$ for GD61 and 0.59$M_\oplus$ for WD0446−255.

## 4. LOOKING FORWARD

### 4.1. *A Large, Uniform Sample of Polluted White Dwarfs*

Polluted white dwarfs have proved to be a powerful way to measure the chemical compositions of exoplanetary material. However, the sample size of polluted white dwarfs is small, and detections of heavy elements are very heterogeneous, as listed in Table 1. Using data from the *Gaia* satellite, Gentile Fusillo et al. (2021a) compiled a sample of 359,000 high-confidence white dwarfs, increasing the number of known white dwarfs tenfold. Large multi-object spectroscopic surveys like the Dark Energy Spectroscopic Instrument (DESI, Cooper et al. 2023), Large Sky Area MultiObject Fibre Spectroscopic Telescope (LAMOST, Guo et al. 2022), SDSS, and the 4-metre Multi-Object Spectroscopic Telescope (4MOST, Chiappini et al. 2019) will identify many more heavily polluted white dwarfs. Dedicated follow-up observations with a high-resolution spectrograph on the current 10-m class telescope or future 30-m class telescopes will return a much larger and more homogeneous sample of polluted white dwarfs with a large number of heavy elements. At the same time, it is important to obtain ultraviolet spectra to constrain the volatile abundances for as many heavily polluted white dwarfs as possible before the end of the *Hubble Space Telescope*.

Measuring the individual isotopic abundances of celestial bodies can yield additional precious information beyond what overall elemental abundances provide. For example, the deuterium-to-hydrogen ratio (D/H) has been measured in many objects in the solar system and is used to trace the origin of Earth's water (Hallis 2017). Unfortunately, this is currently not feasible for polluted white dwarfs. The main challenge is that the isotopic shifts of atomic lines are very small compared to other effects at play in the white dwarf's photosphere. This includes in particular shifts resulting from interactions between the radiating atom and other charged or neutral particles in the dense atmosphere, which cause Stark (for charged particles) and van der Waals (for neutral particles) shift and broadening. Constraining the impact of these interactions on the spectral line shapes remains an area of active research (e.g., Saumon et al. 2022), meaning that potential isotopic shifts are obscured by these other spectral effects.

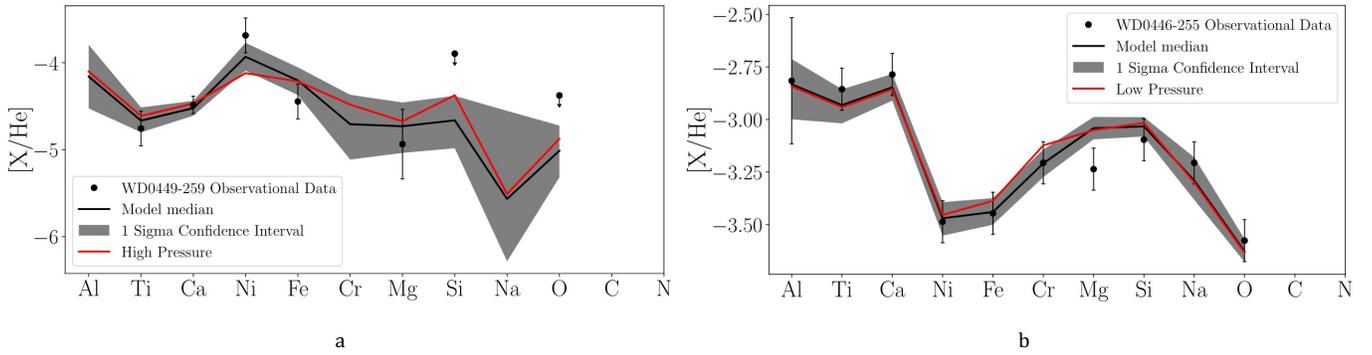

**Figure 12.** Abundance ratios of elements (X) relative to He for WD0449−259 (a) and WD0446−255 (b). Upper limits are shown with arrows. The model with the highest Bayesian evidence is plotted in black with 1$\sigma$ errors as grey shaded regions. The best-fit model for WD0449−259 invokes the accretion of a core-rich fragment of a small differentiated body at low pressures; if core-mantle differentiation of the parent body occurred at high pressures instead, then the abundance pattern would follow the red model. The best-fit model for WD0446−255 invoked the accretion of a mantle-rich fragment of a larger parent body that underwent core-mantle differentiation at a high pressure; the red model shows the abundance pattern for the low pressure case. For WD0446−255, the posterior distribution on the pressure overlaps into the low-pressure region; as can be seen in the plot, it is difficult to differentiate between the two scenarios (black line versus red line) for this white dwarf. More data and better analysis is needed to reduce the uncertainties on the abundances such that these models can be more readily distinguished in the future. Figures adapted from Buchan et al. (2022).

### 4.2. *Improving Abundance Measurements*

#### 4.2.1. *Current treatment of uncertainties*

Abundance ratios published in the literature are ideally assigned an uncertainty $\sigma$ that corresponds to the quadrature sum of two contributions,

$$\sigma^2 (Z_i/\text{He}) = \sigma_{\text{spread}}^2 (Z_i/\text{He}) + \sigma_{T^2_{\text{eff}}}(Z_i/\text{He}). \qquad (2)$$

The first contribution, $\sigma_{\text{spread}}^2 (Z_i/\text{He})$, corresponds to the spread in the elemental abundance inferred using different spectral lines. In general, if a spectrum displays more than one spectral line for a given element, the abundance needed to reproduce each line will differ. This difference stems from uncertainties on the theoretical/experimental line strengths of the lines compiled in line lists, inaccuracies in the model structure, and/or measurement errors. The second contribution $\sigma_{T^2_{\text{eff}}}(Z_i/\text{He})$ has its origin in the uncertainty on the effective temperature (and $\log g$) of the star. Model atmosphere analyses generally yield uncertainties on $T_{\text{eff}}$ of the order of ∼ 5%. By varying the temperature within that range, the inferred elemental abundances will differ. Note that the $T_{\text{eff}}$-induced uncertainties on the abundances of different elements are generally correlated. This implies that the uncertainty on a relative abundance ratio $Z_i/Z_j$ is often smaller than the quadrature sum of the uncertainties on $Z_i/\text{He}$ and $Z_j/\text{He}$, because both $Z_i/\text{He}$ and $Z_j/\text{He}$ move in the same direction as $T_{\text{eff}}$ is changed within its confidence interval. This topic is discussed in more detail in Appendix A of Klein et al. (2021).

Because $Z_i/\text{He}$ can vary by several orders of magnitudes between different stars and elements, it is customary to report abundance measurements and their uncertainties in logarithmic terms. For simplicity, the confidence interval on the abundance is often assumed to be symmetric in this logarithmic space. This assumption can, however, complicate the subsequent analysis of the element-to-element abundance ratios (Doyle et al. 2019; Klein et al. 2021). It should be noted that this choice is somewhat arbitrary and, to date, little effort has been invested in properly characterizing the probability distribution of the measured abundances. Future work that focuses on both better understanding of the uncertainties associated with the abundances of the planetary material, and reducing the uncertainties, will enable drastic improvements in the interpretation of the abundances. Buchan et al. (2022) showed that it is possible to distinguish sizes of bodies from the fragments of core-mantle differentiated bodies using PyllutedWD; however, reduced uncertainties are required to obtain more precise sizes.

#### 4.2.2. *Constitutive physics uncertainties*

Crucially, the current treatment of uncertainties outlined in the previous paragraph largely ignores systematic uncertainties resulting from gaps in our understanding of the constitutive physics of white dwarfs. The atmospheres and envelopes of white dwarfs reach physical conditions unlike those found anywhere on Earth or in most other types of stars. Therefore, modelers are often forced to rely on physics models that have not been directly validated against experimental data. A recent review describes current challenges faced on this front (Saumon et al. 2022); below we briefly discuss some of the most important ones in the context of exoplanetesimal characterization.

**Cool white dwarf atmospheres ($T_{\rm eff} \lesssim 8000$K):** Cool white dwarfs that contain little hydrogen in their outer envelopes have very transparent atmospheres. Helium has the highest ionization potential of all elements, making it very weakly ionized at low temperatures, which results in a very low opacity. As a result, the photospheres of helium-dominated atmospheres are located quite deep and reach densities of up to 1gcm$^{-3}$. Under those liquid-like conditions, many approximations commonly used in stellar atmosphere codes (the ideal gas law, Saha ionization equilibrium, Lorentzian collisional line profiles) break down due to the incessant many-body interactions between atoms, ions and molecules. In recent years, modern quantum chemistry simulation techniques have allowed significant progress on this problem (Kowalski et al. 2007; Kowalski 2014; Allard et al. 2016; Blouin et al. 2017, 2018). Nevertheless, current models still struggle to explain the spectra of many cool white dwarfs (Bergeron et al. 2022; Elms et al. 2022), signaling important gaps in our understanding of the equation of state and opacities under those peculiar conditions. Solving these remaining gaps has become especially pressing since the discovery of old, cool ($T_{\rm eff} \lesssim 5000$K) white dwarfs with Li absorption lines (such as the DZ shown in Figure 2, Kaiser et al. 2021; Hollands et al. 2021), whose interpretation depends on the availability of accurate model atmospheres in this challenging physical regime.

**Diffusion timescales:** A second important source of uncertainty are the diffusion timescales required to translate steady-state photospheric abundances into the composition of the accreting material (as explained in Section 2.3). The diffusion coefficients of the accreted trace elements strongly depend on the ionization state of those same elements at the bottom of the envelope convection zone. The conditions encountered in this region of the star can push ionization models beyond their regime of applicability, especially in the envelopes of cool helium-atmosphere white dwarfs ($T \sim 10^6$ K and $\rho \sim 10^3$ gcm$^{-3}$). The currently available diffusion timescales rely on simple heuristic ionization models that cannot be considered as reliable in this extreme regime (Paquette et al. 1986; Koester et al. 2020). New state-of-the-art simulation techniques can remedy this problem (Heinonen et al. 2020). The outputs of these improved models are not yet available, but they are eagerly awaited as they have been shown to change the inferred abundance ratios by factors of up to three. The uncertainty on the diffusion timescales affects the inferences that are made about the composition of the planetary body accreted by the white dwarf. In order for models like PyllutedWD to better determine exoplanetary composition, the uncertainties on the diffusion times need to be reduced.

**Convective overshoot and thermohaline mixing:** In stellar models, it is common to assume that convection zones end at a well-defined "Schwarzschild boundary" where the fluid becomes stable against convection. In reality, it is well known that such a discontinuity is unphysical (e.g., Zahn 1991; Freytag et al. 1996), and additional convective boundary mixing takes place as, for example, plumes "overshoot" past the formal stability boundary. This extra mixing is important for the interpretation of polluted white dwarfs, as it implies that the accreted material is mixed in a larger mass reservoir (by up to 2-3 orders of magnitude[12], Kupka et al. 2018; Cunningham et al. 2019), thereby implying higher accretion rates and larger accreted bodies. In addition to convective overshoot, the chemical composition gradient induced by the accretion of rocky elements can trigger thermohaline mixing (or "fingering convection", an important mixing process in the Earth's oceans). It is also thought to produce significant extra mixing in white dwarf envelopes (Bauer & Bildsten 2018; Wachlin et al. 2022). 3D hydrodynamic simulations are required to reliably determine the importance of these extra mixing processes (e.g., Tremblay et al. 2015). So far, 3D convective overshoot studies have been limited to a relatively narrow set of temperatures and compositions (Kupka et al. 2018; Cunningham et al. 2019), and no 3D simulations of thermohaline mixing in white dwarfs have been performed.

**Discrepancy between ultraviolet and optical abundances:** A notable discrepancy often emerges when comparing elemental $Z_i$/He abundances derived from ultraviolet observations to those obtained from optical measurements (Gänsicke et al. 2012; Xu et al. 2019; Rogers et al. 2024). The origin of this mismatch remains unclear. Potential culprits include inaccuracies in the model atmosphere structure, vertical composition gradients in the atmosphere, and uncertainties on the line strengths provided in the line lists used when generating synthetic spectra. Any potential solution to this problem has to contend with the fact that this discrepancy is not universal (Klein et al. 2021; Rogers et al. 2024). Fortunately, the relative $Z_i/Z_j$ abundances appear to remain consistent between UV and optical analyses (Rogers et al. 2024), which limits the impact of this problem on planetary composition studies.

---

[12] The effect on relative abundances (e.g., the Ca/Mg abundance ratio) is much smaller.

### 4.3. Connecting Exoplanetary Compositions with the Solar System

Stronger collaboration between the polluted white dwarf community and the geochemistry community is urgently needed to interpret the abundance measurements. For example, Doyle et al. (2019, 2020) calculated the oxygen fugacity for a sample of polluted white dwarfs and found most extrasolar rocky bodies differentiated under oxidizing conditions compared to the solar gas. Buchan et al. (2022) proposed a new way to constrain the size of the polluting bodies from the abundances of Ni, Cr, and Si, and this size can be constrained using PyllutedWD. However, this relies on accurate and precise metal-silicate partitioning coefficients which remain uncertain for high pressures. Therefore, to improve the understanding of core-mantle differentiation and the size of the exoplanetary bodies that core-mantle fragments come from, further experiments on the partitioning need to be conducted. These endeavors are particularly welcomed to put planetary abundances measured from polluted white dwarfs in the context of the solar system and the more general planet formation and evolution scenario.

### 4.4. Dust & Gas Composition

Due to the strong gravitational pull from the white dwarf, extrasolar planetesimals are tidally disrupted, making it easier to study the interior bulk compositions in comparison to indirectly inferring it. Here, we describe two other promising ways to measure the compositions of exoplanetary material using white dwarf planetary systems.

**Composition of circumstellar dust:** Infrared emission from dusty material within the tidal disruption radius of white dwarfs has now been observed in over one hundred systems (Lai et al. 2021). Infrared spectroscopic observations of these dust disks reveal the mineralogy of this material and provide a way to directly probe exoplanetary geology. The Spitzer Infrared Spectrograph (IRS) observed the brightest polluted white dwarfs with dust disks; all spectra show prominent silicate features around $10\mu$m (Jura et al. 2009). Amorphous versus crystalline silicates can be readily identified in the spectra as crystalline silicates create sharp peaks in the $10\mu$m feature, whereas amorphous silicates show smooth, featureless $10\mu$m features. The mid-infrared spectrum of G29–38 is dominated by amorphous silicate (Reach et al. 2009), whereas the others show evidence for crystalline features related to silicate minerals. The enhanced sensitivity and resolution provided by the *James Webb Space Telescope* (*JWST*) will enable the number of white dwarf disks with mineralogy measurements to increase tenfold, as well as more in-depth studies of the mineralogy. For example, the $10\mu$m crystalline silicate emission feature is distinguishable dependent on whether the silicate mineralogy is olivine or pyroxene dominated, with the dominant feature shifting from $\sim 9\mu$m for pyroxene to $\sim 11\mu$m for olivine. This affects the water storage capacity of mantles and its ability to sustain plate tectonics, crucial for questions regarding habitability (e.g., Kelley et al. 2010; Lambart et al. 2016; Hinkel & Unterborn 2018; Wang et al. 2022). With a number of *JWST* proposals accepted that will study the dust mineralogy of polluted white dwarf disks, over the next few years our understanding of the mineralogy of exoplanetary material will be revolutionized.

**Composition of circumstellar gas:** A small fraction of white dwarfs with circumstellar dust also show circumstellar gas, mostly as double-peaked emission features (e.g., Gänsicke et al. 2006) but some as additional absorption features (e.g., Debes et al. 2012). Thanks to dedicated follow-up studies from the new *Gaia* white dwarfs, the number of white dwarfs with circumstellar gas detections has increased significantly (Melis et al. 2020; Dennihy et al. 2020; Gentile Fusillo et al. 2021b). WD 1145+017 has the most elements detected in circumstellar gas, including Ca, Mg, Ti, Cr, Mn, Fe, and Ni (Xu et al. 2016). There has been significant progress on the modelling front as well to understand the properties of the circumstellar gas (Gänsicke et al. 2019; Fortin-Archambault et al. 2020; Steele et al. 2021). Measuring the chemical compositions of the circumstellar gas around white dwarfs will be a powerful way to constrain the composition of the exoplanetary material and contrast with the measurements from the polluted atmospheres.

## 5. CONCLUSIONS

Spectroscopic observations of polluted white dwarfs measure the bulk compositions of extrasolar planetary material, which is not possible with any other technique. The number of heavily polluted white dwarfs has increased significantly over the past decade with many new systems having unique abundance ratios with no solar system analog. Looking to the future, improvements on white dwarf model atmospheres, a larger uniform sample of polluted white dwarfs, and a stronger connection between the white dwarf community and the geochemistry community are needed to interpret these abundance measurements. Polluted white dwarf studies are essential for assessing a planet's overall habitability from on-going and future missions such as JWST and HWO.

*Acknowledgement* The authors thank Rhian Jones and E. D. Young for discussions of comparing abundance measurements from polluted white dwarfs with those of the meteorites and Andy Buchan for discussions related to PyllutedWD. S. Xu is supported by the international Gemini Observatory, a program of NSF's NOIRLab, which is managed by the Association of Universities for Research in Astronomy (AURA) under a cooperative agreement with the National Science Foundation on behalf of the Gemini partnership of Argentina, Brazil, Canada, Chile, the Republic of Korea, and the United States of America. L. K Rogers acknowledges support of a Royal Society University Research Fellowship, URF\R1\211421 and an ESA Co-Sponsored Research Agreement No. 4000138341/22/NL/GLC/my = Tracing the Geology of Exoplanets. S. Blouin is a Banting Postdoctoral Fellow and a CITA National Fellow, supported by the Natural Sciences and Engineering Research Council of Canada (NSERC).

The research shown here acknowledges use of the Hypatia Catalog Database, an online compilation of stellar abundance data as described in Hinkel et al. (2014, AJ, 148, 54), which was supported by NASA's Nexus for Exoplanet System Science (NExSS) research coordination network and the Vanderbilt Initiative in Data-Intensive Astrophysics (VIDA).